\documentclass[aps,prl,twocolumn, superscriptaddress, amsmath]{revtex4-1}
\usepackage{natbib}
\usepackage{graphicx}
\usepackage{bm}
\usepackage{color,soul}
\usepackage{hyperref}% add hypertext capabilities
\usepackage{float}
\usepackage{csquotes}
\usepackage{color,soul}
\usepackage{hyperref}% add hypertext capabilities
\hypersetup{colorlinks=true, urlcolor=blue, citecolor=blue}
\usepackage{longtable}
\usepackage{tabularx}
\usepackage{adjustbox}
\usepackage{braket}

\setcitestyle{journalcolor= blue}

\usepackage{babel}

\begin{document}
% \title{Stacking-order independent anisotropic in-plane lattice thermal conductivity in ReS$_{2}$}
\title{Phonon Band Center: A Robust Descriptor to Capture Anharmonicity}
\author{Madhubanti Mukherjee}
\author{Ashutosh Srivastava}
\author{Abhishek Kumar Singh}
\email {abhishek@iisc.ac.in}
\affiliation{Materials Research Centre, Indian Institute of Science, Bangalore 560012, India}

\date{\today}
\begin{abstract}
Understanding anharmonicity is crucial for designing materials with desired lattice thermal conductivity. Designing a material descriptor that effectively captures anharmonicity while being cost-effective remains a significant challenge. This work proposes a simple metric that helps explain the diversity in lattice thermal conductivity ($\kappa_l$) among materials by quantifying their anharmonic effects. This descriptor ``phonon band center" (PBC) encapsulates the critical factors associated with the physics of phonon scattering, revealing a simple inverse relationship with the Gr$\ddot{u}$neisen parameter, the response of phonons with changing volume, and strong correlation with lattice thermal conductivity. This metric has been established using the chalcopyrite class of materials and subsequently validated across various classes of materials using experimental $\kappa_l$. Our approach effectively differentiates materials based on PBC, thereby streamlining the identification of candidates with desirable $\kappa_l$.
      
\end{abstract}
   
\maketitle

\begin{figure*}[!htbp]
\centering
\includegraphics[width=1.0\linewidth]{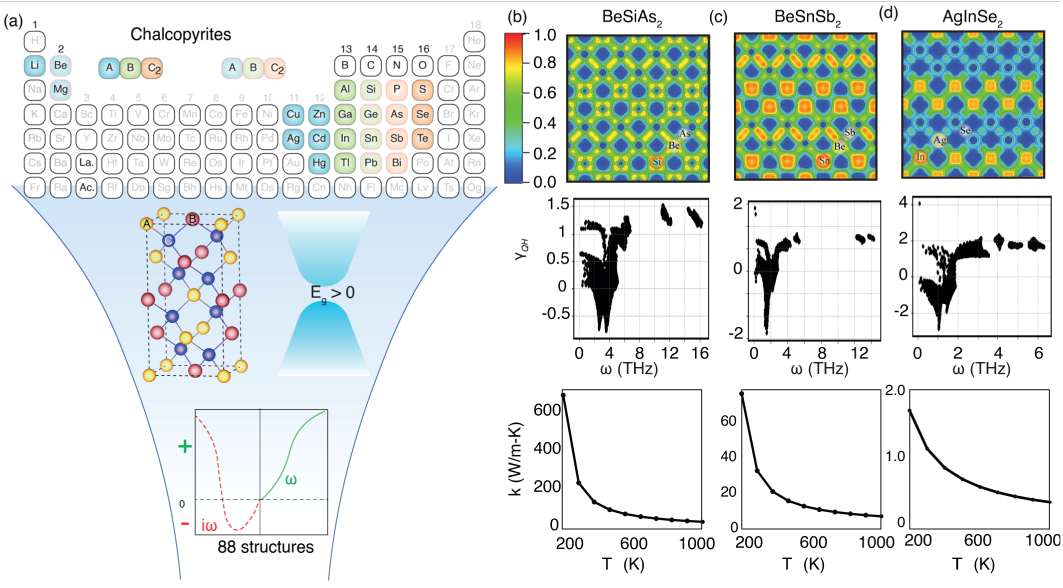}
\caption{(a) High throughput workflow, column-wise calculated electron localization function (ELF), variation of Gr$\ddot{u}$neisen parameter ($\gamma_{QH}$) as a function of frequency and lattice thermal conductivity ($\kappa$) for three representative compounds having (b) ultrahigh (BeSiAs$_{2}$), (c) medium (BeSnSb$_{2}$), and (d) ultralow (AgInSe$_{2}$) range of thermal conductivity.}
\label{fig:1}
\end{figure*}

\section{Introduction}
Anharmonic effects are inherent factors that play a critical role in determining the thermal transport of materials. Such effects arise from higher-order phonon-phonon interactions, and as the effects start to dominate, the ability of phonons to carry heat decreases, resulting in lower lattice thermal conductivity ($\kappa_l$) (such as SnSe~\cite{zhao2014ultralow}). Discovering new materials with exceptionally high or low $\kappa_l$ is essential to advance technologies such as thermoelectrics, where thermal insulation is vital, or electronics, where efficient heat dissipation is crucial. Although there has been substantial progress in computational methods for predicting $\kappa_l$, \textit{ab-initio} approaches remain computationally expensive for large-scale exploration of material space.~\cite{carrete2014finding,seko2015prediction,purcell2023accelerating} 

In this context, having an a priori assessment of a material's anharmonic characteristics can be highly advantageous. Such insights would guide the search for materials with targeted thermal properties more efficiently, saving time and resources. Hence, it is important to categorize materials according to the extent of their anharmonicity systematically. Identifying parameters or descriptors that can effectively correlate anharmonicity with $\kappa_l$ without the need for computationally expensive calculations could greatly accelerate material discovery.

%why we need PBC is clear. But what are the advantages that it possesses over previously proposed descriptors? One can be it is independent of the class/type of structure/material we consider. 

Several approaches have been developed to capture anharmonic effects, including expensive anharmonic lattice-dynamics calculations, non-perturbative \textit{ab-initio} molecular dynamics (AIMD), and using perturbative approaches~\cite{seko2015prediction, carbogno2017ab, marcolongo2016microscopic, broido2005lattice, esfarjani2011heat, tadano2014anharmonic, hellman2014phonon, tadano2015self, feng2017four, ravichandran2018unified, simoncelli2019unified}. Although these approaches have higher accuracy, increasing computational resources and time with the system size hinders their applicability to complex systems. Within the quasi-harmonic approximation (QHA), the anharmonicity in materials can be captured through Gr$\ddot{u}$neisen parameter by calculating changes in the phonon frequency with volume. However, the temperature dependence of Gr$\ddot{u}$neisen parameter can only be captured by considering third-order IFCs. However, obtaining accurate IFCs of any order still requires multiple computational calculations. Therefore, a simple and robust quantitative metric that can efficiently capture anharmonicity and serve as a reliable indicator of lattice thermal conductivity across a wide chemical space is required.

In this article, we devise a potential descriptor, "Phonon Band Center" (PBC), based on first-principles calculations, which essentially captures the frequency-weighted phonon density of states of any given material. We construct this descriptor stepwise using a high-throughput screening on a dataset of 236 ternary chalcopyrites. The analysis of phonon dispersions led to a set of 88 dynamically stable compounds, for which the Gr$\ddot{u}$neisen parameters were calculated. By extracting information on phonon interactions, Gr$\ddot{u}$neisen parameter, the contribution of each relevant physical factor, such as mass, volume, atomic vibrations, and bonding nature, is analyzed and quantified. This renders the parameter "PBC" as a potential descriptor exhibiting an inverse relationship with Gr$\ddot{u}$neisen parameter, thereby capturing anharmonic characteristics of the chalcopyrites. In addition, this descriptor can categorize a diverse set of materials with high and low $\kappa_l$, which is validated against their experimental $\kappa_l$ values at 300K, implying a greater advantage.

\section{Computational Methodology}
The calculations are carried out by using first-principles density functional theory (DFT), as implemented in the Vienna \textit{ab-initio} simulation package (VASP)~\cite{kresse1996efficient, kresse1996efficiency}. The electronic exchange and correlation potential has been approximated by the Perdew-Burke-Ernzerhof (PBE) under the generalized gradient approximation (GGA)~\cite{perdew1996generalized, blochl1994projector}. The ion-electron interactions were represented by the projector augmented wave (PAW) potentials~\cite{kresse1999ultrasoft}. The plane wave energy cut-off was set to 1.3 times the default value in the PAW pseudopotential. A strict energy convergence criterion of 10$^{-6}$ eV was used. The phonon dispersion and Gr$\ddot{u}$neisen parameter (within quasi-harmonic approximation) have been calculated using the supercell approach, as implemented in the PHONOPY~\cite{togo2015first}. The temperature-dependent Gr$\ddot{u}$neisen parameter has been calculated by solving the semiclassical phonon Boltzmann transport equation, as implemented in the ShengBTE code~\cite{li2014shengbte}. Solving PBTE requires second-order harmonic and third-order anharmonic IFCs obtained from Phonopy and the third-order code~\cite{togo2015first, li2014shengbte}. For anharmonic IFCs, the interactions up to the third nearest neighbor have been considered. Well-converged supercell sizes of 4$\times$4$\times$2 and 2$\times$2$\times$1 were used for the harmonic and anharmonic IFCs calculations, respectively. A k-grid of 4$\times$4$\times$2 and an energy cutoff of 600 eV with a strict energy convergence criterion of 10$^{-8}$ eV were used to obtain accurate forces. The Born effective charges were calculated by considering long-range electrostatic interactions using density functional perturbation theory (DFPT).

\section{Results and Discussions}
\subsection{Database and high-throughput strategy}

We started with a total of 236 ternary chalcopyrites with I-III-VI$_{2}$ and II-IV-V$_{2}$ semiconductors containing atomic and vibrational properties. These diamond-like structures provide a valuable platform for exploring a wide compositional space through isoelectronic substitution, leading to elemental and chemical diversity and significantly influencing phononic properties. Using the high-throughput workflow (Figure~\ref{fig:1} (a)) on the chalcopyrite compounds, we find 88 dynamically stable chalcopyrites. The phonon dispersions for all these compounds are shown in Figures S1 and S2. The corresponding Gr$\ddot{u}$neisen parameters ($\gamma_{QH}$) representing the anharmonicity in these compounds, calculated within a QHA, are shown in Figures S3 and S4. The temperature dependence of the Gr$\ddot{u}$neisen parameter has been considered by obtaining third-order interactions as implemented in ShengBTE.\cite{li2014shengbte} Anharmonicity at room temperature in these materials has been quantified by calculating the total $\gamma_{300}$ obtained as a weighted sum of the mode contributions at 300K~\cite{maradudin1962thermal, wu2011quasiharmonic, cuffari2020calculation, bjerg2014modeling, xie2019scattering}. 

Three representative materials spanning several orders of magnitude in $\kappa_l$ have been chosen to analyze the phonon characteristics. For example, BeSiAs$_{2}$ with higher $\kappa_{l}$ has phonon modes spanning a larger frequency range as compared to that of BeSnSb$_{2}$, and AgInSe$_{2}$, having an order of magnitude lower $\kappa_l$ than BeSiAs$_2$, as shown in Figures~\ref{fig:1}(b) to (d). Interestingly, despite having a nearly similar frequency range of phonon modes, slightly larger $\gamma_{QH}$ in BeSnSb$_{2}$ leads to a significant reduction in $\kappa_l$, compared to BeSiAs$_{2}$. Such a substantial impact of anharmonic strength on thermal conductivity demands a systematic understanding of the anharmonic characteristics of materials.

\subsection{Underlying physics of anharmonicity}

Vibrational characteristics and interactions between phonons in materials are usually determined by calculating phonon frequencies, phonon group velocities, and harmonic and anharmonic IFCs.~\cite{mukherjee2022recent} Within the harmonic approximation, the lattice dynamics properties are determined by the following Hamiltonian~\cite{huang1954dynamical, dove1993introduction}:
\begin{equation}
H = \sum_{I}\frac{P_{I}^{2}}{2M_{I}} + \frac{1}{2}\sum_{i,j,\alpha,\beta}\phi_{\alpha,\beta}^{I,J} \Delta R_{I}^{\alpha} \Delta R_{I}^{\beta}
\end{equation}
where the first term is nuclei kinetic energy, $\Delta R_{I}$ are the atomic displacements, and $\phi_{\alpha,\beta}^{I,J}$ is the second order IFCs. Due to the straightforward and simple computations involved in determining $\phi_{\alpha,\beta}^{I, J}$, the harmonic approximation has become a successful and widely used approach to understand phonon behaviors ~\cite{plata2017efficient, togo2015first, parlinski1997first}. Anharmonicity is the key factor responsible for making the lattice thermal conductivity finite in real materials~\cite {leibfried1961theory, klemens1958thermal}. In general, a higher anharmonicity leads to a lower $\kappa_{l}$ in the materials. This emphasizes the need for a more comprehensive understanding of metrics that quantify anharmonicity and can be easily obtained.

At a given temperature, mode Gr$\ddot{u}$neisen parameter ($\gamma_{m}$) can be written as~\cite{barron1974dynamical, fabian1997thermal, broido2005lattice}:
\begin{equation}
\gamma_{m} = - \frac{1}{6\omega^{2}_{m}} \sum_{ijk \alpha \beta \gamma} \frac{\epsilon^{m*}_{i\alpha}\epsilon^{m}_{j\beta}}{\sqrt{M_{i}M_{j}}} r^{\gamma}_{k} \psi^{\alpha\beta\gamma}_{ijk} e^{iq.r_{j}}
\end{equation}
where $\psi^{\alpha\beta\gamma}_{ijk}$ are the coefficients of the quadratic terms in the Taylor expansion of potential energy. $\epsilon^m_{i\alpha}$ is the vibrational eigenvector for atom i. $M_{i}$ is the mass of atom i and r$_{i}$ is the vector describing its position. The mode Gr$\ddot{u}$neisen parameter in equation 2 can further be decomposed into small blocks to isolate and quantify the contributions of various relevant physical factors governing anharmonicity. 

Since $\gamma_{m}$ is directly related to cubic anharmonic IFCs, it is essential to analyze the dependence of $\psi^{\alpha\beta\gamma}_{ijk}$ on the atomic level descriptions. Third-order IFCs have a direct dependence on the three-phonon scattering strength, which encapsulates the effect of atomic type, structural geometry, electronic density, and phonon dynamics and is therefore an intrinsic characteristic of the material~\cite{wallace1972thermodynamics}. It is described as follows,
\begin{equation}
\begin{split}
\phi_{\lambda \lambda^{'} \lambda^{''}} \propto \frac{1}{\sqrt{\omega_{\lambda}\omega_{\lambda^{'}}\omega_{\lambda^{''}}}} \sum_{kk^{'}k^{''}} \frac{1}{\sqrt{M_{k}{M_{k^{'}}{M_{k^{''}}}}}}  \\
\times \sum_{\alpha \beta \gamma} \epsilon^{\alpha}_{\lambda}(r_{k}) \epsilon^{\beta}_{\lambda^{'}}(r_{k^{'}}) \epsilon^{\gamma}_{\lambda^{''}}(r_{k^{''}}) \\
\times \sum_{ll^{'}l^{''}} e^{iq.r_{kl}} e^{iq^{'}.r_{k^{'}l^{'}}} e^{iq^{''}.r_{k^{''}l^{''}}} \psi^{\alpha\beta\gamma}_{ijk} (r_{kl}, r_{k^{'}l^{'}}, r_{k^{''}l^{''}})
\end{split}
\end{equation}
where $\omega_{\lambda}$ is the eigenfrequency corresponding to the mode $\lambda$, $M_{k}$ is the mass of the k$^{th}$ atom, r$_{kl}$ (r$_{k^{'}l^{'}}$, r$_{k^{''}l^{''}}$) are the position of the k$^{th}$ (k$^{'th}$, k$^{''th}$) atom in the l-th (l$^{'th}$, l$^{''th}$) cell, $\epsilon^{\alpha}_{\lambda}(r_{k})$ is the $\alpha^{th}$ cartesian component of the eigenvector associated to the mode $\lambda$ and to the k$^{th}$ atom. This equation suggests that IFCs $\psi^{\alpha\beta\gamma}_{ijk}$ depend on various attributes such as the type of atoms and the environment in which atoms are embedded, the distributions of electronic density, the degree of bond ionicity, and the vibrational properties of the lattice. The cubic anharmonic terms or the anharmonic phonon scattering are thus determined by the eigenvectors and eigenfrequencies derived from the above factors. Therefore, anharmonicity strongly depends on various inherent aspects of the system via the modulus of multiple $\phi_{\lambda \lambda^{'} \lambda^{''}}$ terms. Designing a simple and general descriptor is crucial to capture these parameters into a unifying framework.

\subsection{Physical parameters to descriptor mapping}

To understand the effect of the parameters mentioned above on anharmonicity, we analyze the phonon dispersion, the nature of bonding, the atomic types, and the vibrations of the atoms. Considering the three representative materials (BeSiAs$_{2}$, BeSnSb$_{2}$, and AgInSe$_{2}$), a larger $\gamma_{QH}$ has been observed for the one with the lowest maximum phonon frequency ($\omega_{max}$), as shown in Figures S5 (a) to (c). As mentioned previously, BeSiAs$_{2}$ and BeSnSb$_{2}$ possess nearly similar range of $\omega_{max}$, but have a one order of magnitude difference in $\kappa_l$, due to the variation in $\gamma_{QH}$. The larger $\gamma_{QH}$ in BeSnSb$_{2}$ can be attributed to the softening of the phonon mode in the low-frequency range, as shown in Figure S5 (b), leading to a strong hybridization of the acoustic and optical modes and enhanced anharmonicity. Furthermore, the $\gamma_{QH}$ spread throughout the frequency range exhibits distinct variations, previously shown as a potential descriptor of the prediction $\kappa_l$~\cite{juneja2019coupling}. The span of Gr$\ddot{u}$neisen parameter across the frequency ranges must be significantly influenced by the available phonon states for scattering in the corresponding frequency region. The distribution of phonon modes in different frequency regions is determined by several physical parameters such as the mass of the atoms, geometric arrangements, and the bonding nature between the atoms. On the basis of these parameters, atoms contribute to different frequency ranges. The atomic contribution to the phonon modes is captured by the projected phonon density of states (PDOS). A closer look at the PDOS of BeSiAs$_{2}$, BeSnSb$_{2}$, and AgInSe$_{2}$, as shown in Figures S5 (a)-(c), reveals the presence of a large number of low-frequency phonon modes arising from the large amplitude of vibrations of different atoms. play a significant role in deciding the anharmonicity in the system. For instance, unlike BeSiAs$_2$ and BeSnSb$_{2}$, PDOS of AgInSe$_{2}$ exhibits large and collective contributions from Ag, In, and Se in the low-frequency ranges and a more extensive $\gamma_{QH}$. The mass difference between the constituent atoms commonly determines the contribution of specific atoms in the low-frequency region. However, the overlapped states in PDOS of AgInSe$_{2}$ suggest that the atomic vibrations are coupled despite the mass difference. Vibrational modes are therefore the result of the contributions of each ion of a distinct atomic type, depending on the bonding environment. The coexistence of mass difference and anisotropy in chemical bonds therefore leads to a significant and collective contribution to the vibration for AgInSe$_{2}$, which is not the case for BeSiAs$_{2}$. 

We calculate the electron localization function (ELF) to assess the bonding variation. ELF can have values of 1 to 0, where ELF = 1, and 0 represents the complete localization and delocalization of the electrons, respectively~\cite{savin1997elf, perdew1981self, becke1990simple}. As shown in Figure~\ref{fig:1}(b), ELF values remain 0.6-0.8 between Be-Si and Be-As, indicating an isotropic and stronger bonding in the system. At the same time, a delocalization around the Be atoms is observed. In addition to the mass dependence, strongly bonded Be atom states are shifted to the upper phonon frequency. A similar trend has been observed for BeSnSb$_{2}$, where, despite a localization around Sn and Sb, the Be-Sn bonds show an ELF of 0.4-0.6 (Figure~\ref{fig:1}(c)), while the ELF values between Be-Sb remain similar to those of Be-As. A slight reduction in the relative bond strength and comparable mass of Sb and Sn thus leads to more significant contributions in low-frequency phonon modes compared to that of BeSiAs$_{2}$. The importance of bonding is most prominent for the case of AgInSe$_{2}$ (Figure~\ref{fig:1}(d)), where Ag-Se bonds have an ELF ranging from 0.0-0.2, and Ag-In bonds share a mixed covalent bonding with ELF value 0.4. The weak bonding of Se atoms with both Ag and In thus leads to significant amplitude vibrations in the frequency range of 0-2 THz. Furthermore, the delocalization around the Ag results in the most crucial contribution of Ag vibrations in low-frequency phonon modes. A more significant number of states in PDOS ensures the phonon softening and hybridization, and, in turn, the anharmonicity in the materials. Therefore, atomic mass and chemical bonding between the atoms broadly capture the $\gamma_{QH}$. 

An effort to quantify the contributions from these physical factors to the anharmonicity renders phonon density of states as a standard attribute capturing all these effects. The phonon states involve complex functions of all the components of the IFCs tensors associated with all the atomic pairs present in the system. The smallest unit that describes this dynamic interaction is the average phonon energy the atomic pairs produce. To obtain the most straightforward metric to represent the dynamical interactions, ``phonon band center (PBC)" has been defined, which is phonon frequencies weighted by the phonon density of states. PBC essentially quantifies the average vibrational frequency of any given material, with a mathematical form as follows:
\begin{equation}
PBC = \frac{\int \omega \times DOS (\omega) d\omega}{\int DOS (\omega) d\omega}
\end{equation}
PBC can be understood by considering PDOS with varying numbers of states in different phonon frequency ranges. The phonon band center is thus determined according to the weightage of the vibrational modes distributed across the phonon dispersion. Materials with equally distributed phonon modes over the phonon frequency ranges will have PBC in the middle. Whereas materials having phonon modes biased in one frequency region, the position of the PBC will be accordingly. A large population of phonon modes in the low-frequency areas leads to phonon softening, which arises from heavier masses or weak bonding and significantly influences materials' anharmonicity. Softening of phonon modes will, therefore, result in a lower value for PBC. In this manner, PBC acts as the centroid of the PDOS, capturing the information regarding bonding, phonon softening, mass, and volume dependence of the atomic vibrations. This represents the potential of PBC to be considered as a metric to estimate the strength of anharmonicity in a material.

Potential energy surfaces of materials are notably important to analyze materials' anharmonicity. Typically, within harmonic approximation, potential energy varies as the square of the displacement. The deviations from this ideal harmonic behavior refer to anharmonicity, where the potential energy surface deviates from a simple parabolic shape, becoming flatter or more complex at larger displacements. Interestingly, this trend has also been captured by the phonon band center. The potential energy surfaces for individual atoms of the three representative compounds are shown in Figure S6. The potential energy surfaces exhibit a significantly shallow potential well for Ag atoms in AgInSe$_{2}$. In contrast, the BeSnSb$_{2}$ and BeSiAs$_{2}$ show relatively deeper potential well. Furthermore, we calculated the elemental contribution to PBC (see Table S1), showing that the individual contribution of atoms in PBC values excellently captures the anharmonicity arising from the respective atoms. For instance, Ag atoms contributing to the shallow potential energy surface in AgInSe$_{2}$ exhibit the lowest PBC value among the three elements present in the system. 

\begin{figure*}[!htbp]
\centering
\includegraphics[width=1.0\linewidth]{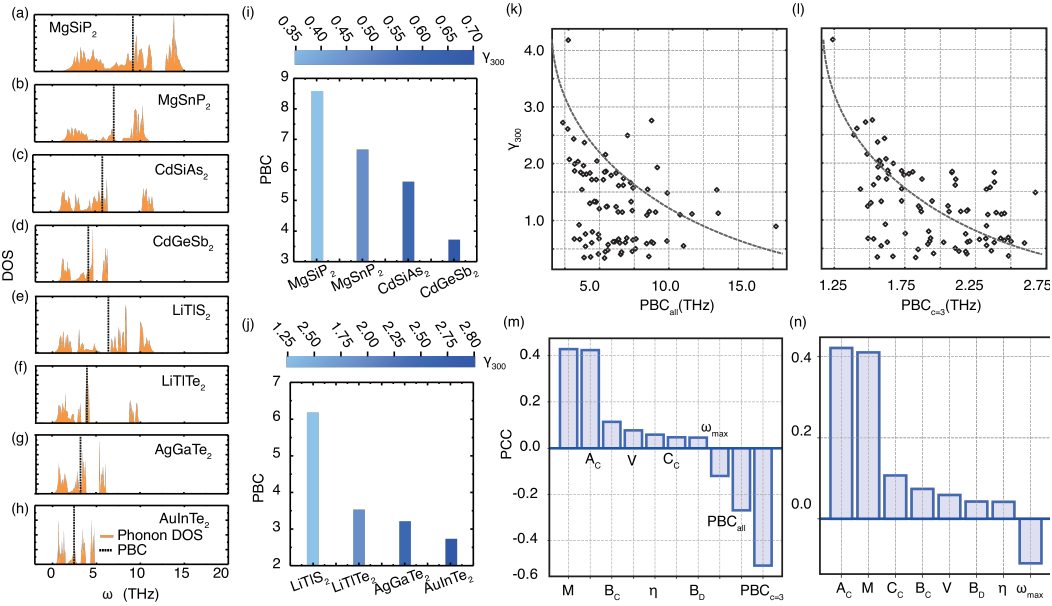}
\caption{Phonon density of states (PDOS) plot of (a)-(d) representative compounds from group II-IV-V$_{2}$, and (e)-(h) representative compounds from group I-III-VI$_{2}$ chalcopyrites. The black dotted
line indicates the position of the total phonon band center (PBC). Corresponding calculated PBC values for the compounds from (i) II-IV-V$_{2}$, and (j) I-III-VI$_{2}$ chalcopyrites, where the color bar represents the value of $\gamma_{300}$. Variation of $\gamma_{300}$ with calculated phonon band center for (k, m) entire frequency range, and (l, n) considering a cut-off frequency range of 3 THz.}
 \label{fig:2}
\end{figure*}

To further evaluate the significance and robustness of PBC, we next analyze the PBCs and the temperature-dependent Gr$\ddot{u}$neisen parameters ($\gamma_{300}$) for a few compounds from II-IV-V$_{2}$, and I-III-VI$_{2}$ chalcopyrite classes. PBC varies in a wide range from 3.5-8.5 for II-IV-V$_{2}$ chalcopyrites, as shown in Figures~\ref{fig:2}(a)-(d). MgSiP$_{2}$ shows the highest PBC with a higher maximum phonon frequency. In contrast, CdGeSb$_{2}$ has a much lower PBC and significantly lower $\omega_{max}$. This difference has been reflected in the $\gamma_{300}$, exhibiting significantly enhanced anharmonicity in CdGeSb$_{2}$, as shown in Figure~\ref{fig:2}(i). Interestingly, MgSnP$_{2}$ and CdSiAs$_{2}$ exhibit similar maximum phonon frequency; however, there is a notable difference in PBC values due to variation in phonon distribution in the low-frequency region. This results into a larger $\gamma_{300}$ for CdSiAs$_{2}$, owing to a lower PBC, than that of MgSnP$_{2}$. Therefore, the distribution of phonon modes becomes significantly important for capturing the anharmonicity. For instance, larger phonon frequency and an equal distribution of phonon modes arise from the stiff lattice and isotropic bonding. The softening of phonon modes in low-frequency regions commonly occurs in soft lattices with loosely packed atoms having anisotropic bond strength. A similar trend in PBC and $\gamma_{300}$ has been observed for I-III-VI$_{2}$ chalcopyrites, as shown in Figures~\ref{fig:2}(e)-(h), and (j). However, an overall narrower range in PBC for I-III-VI$_{2}$ chalcopyrites leads to a larger $\gamma_{300}$, as compared to that of II-IV-V$_{2}$ chalcopyrites. The universally lower PBC of I-III-VI$_{2}$ can be attributed to anisotropic bonding and more phonon states at low frequencies. PBC thus effectively encapsulates the fundamental physics governing anharmonicity.

Next, the comprehensive dataset has been analyzed to validate and generalize the insights obtained from the variation of $\gamma_{300}$ with the calculated PBC for the representative materials. Figure~\ref{fig:2}(k) shows an decreasing trend of $\gamma_{300}$, with increasing PBC values. An inverse relation between PBC and $\gamma$ indicates that strongly anharmonic materials would possess localized low-frequency phonon modes arising from mass, bonding, and atomic arrangements, which lower the PBC. Compared to the entire frequency range, low-lying phonon modes predominantly determine the anharmonic nature of a material. Therefore, the PBC within a frequency range of 3 THz could represent the anharmonicity better. As was foreseeable, Figure~\ref{fig:2}(l) exhibits a more prominent inverse dependence of $\gamma_{300}$ with PBC$_{c=3}$. A combination of elemental, structural, and lattice dynamics descriptors such as maximum phonon frequency $\omega_{max}$, mass (M), and volume of the unit cell (V) are widely used for the prediction of lattice thermal conductivity with unprecedented accuracy~\cite{seko2015prediction, chen2019machine, wang2020identification, juneja2020guided}. Since $\kappa_l$ has a direct correlation with anharmonicity, we checked the statistical correlation between $\gamma_{300}$ and the descriptors, including M, V, $\omega_{max}$, average bond distance (B$_{D}$) (an alternative measure for nature of bonding), specific atomic contribution to phonon modes (A$_{c}$, B$_{c}$, C$_{c}$), tetragonal distortion ($\eta$) and PBC (particularly up to 3 THz)~\cite{mukherjee2018high}. As shown in Figures~\ref{fig:2}(m) and (n), the PBC up to 3 THz (PBC$_{c=3}$) possesses the maximum correlation of $-$0.6 with $\gamma$ as compared to other descriptors, including the PBC for the entire frequency range. Therefore, PBC$_{c=3}$ has been defined as the simplest effective descriptor for predicting the degree of anharmonicity in a material without explicitly calculating third-order IFCs. 

\subsection{Universality of phonon band center}
\begin{figure}[htb!]
\centering
\includegraphics[width=1.0\columnwidth]{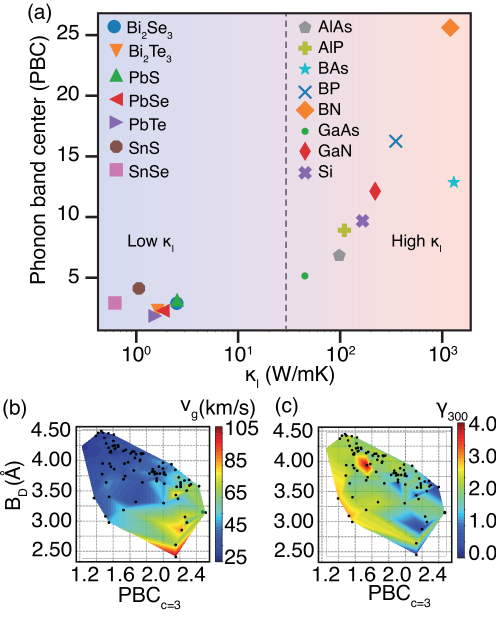}
\caption{(a) Experimentally measured lattice thermal conductivity ($\kappa_l$) values of Bi$_{2}$Se$_{3}$, Bi$_{2}$Te$_{3}$, PbS, PbSe, PbTe, SnS, SnSe, AlAs, AlP, BAs, BP, BN, GaAs, GaN, and Si ~\cite{wu2023experimental,zhao2014ultralow,shinde2006high,sze2021physics,karim1993characterization,kang2018experimental,morelli2006high,jezowski2003thermal,fournier2018straightforward,bessas2012lattice,pei2012electrical,banik2016origin} at 300 K and corresponding calculated phonon band center (PBC) values calculated using \textit{first-principles}. Contour plot showing the variation of PBC$_{c=3}$ as a function of average bond distance (B$_{D}$), where the color bars represent the magnitude of (a) $\nu_{g}$, and (b) $\gamma_{300}$.}
 \label{fig:3}
\end{figure}

To corroborate the effectiveness of PBC as a universal descriptor for a vast search space, we extended our analysis to additional material classes by compiling experimental values of the thermal conductivity ($\kappa_{l}$) at room temperature for a set of materials known for their either high or low $\kappa_{l}$ (Table S2). Subsequently, we calculated the phonon spectra and determined the value of PBC for these materials to examine the relationship between these parameters thoroughly. Figure~\ref{fig:3}(a) illustrates a distinct correlation between PBC and $\kappa_{l}$ of these materials. Notably, materials with a higher $\kappa_l$, such as AlN, BN, and BAs, consistently demonstrate higher PBC values (as highlighted by the light red region). In contrast, materials with lower $\kappa_l$, such as PbTe and SnSe, are marked by significantly reduced PBC values (as highlighted by the light blue region). The calculated $\gamma_{QH}$ (plotted in Figure S7) of these materials are also in agreement with the fact that PBC captures the anharmonicity accurately. Materials having higher $\gamma_{QH}$ values fall in lower PBC regimes of Figure~\ref{fig:3}(a). This underscores the utility of PBC in classifying high and low thermal conductivity materials merely through its calculation, thereby illustrating the robustness of PBC in capturing anharmonicity and filtering high/low $\kappa_{l}$ materials, independent of the materials class and structure. Additionally, this descriptor offers experimentalists the advantage of obtaining meaningful insights before conducting rigorous experimental measurements.

As obtaining PBC requires only second-order IFCs, capturing the harmonic properties, such as group velocity, can be useful. According to classical kinetic theory, lattice thermal conductivity has a direct dependence on group velocity ($\nu_{g}$), implying a similar inverse dependence of $\nu_{g}$ with anharmonicity. Since the bonding information is best quantified by average bond distance, we plotted PBC$_{c=3}$ against B$_{D}$ as shown in Figures~\ref{fig:3}(b) and (c), where the magnitude of $\nu_{g}$, and $\gamma_{300}$ are represented by the color bar, respectively. Smaller PBC, and larger B$_{D}$ lead to a low $\nu_{g}$, and high $\gamma_{300}$. This is because larger B$_{D}$ implies weaker bonding between atoms, resulting in large amplitude vibrations of the weakly bonded atoms in a low-frequency regime. A large population of phonon states in a lower frequency range implies a low PBC value. Localized phonon states emerging from the collective contribution of atoms lead to flat phonon modes, which reduce the propagation velocity of phonons or the group velocity. On the other hand, many phonon states enhance the phonon scattering channel, yielding strong anharmonicity. Therefore, PBC$_{c=3}$ connects the harmonic and anharmonic regimes without explicitly calculating the expensive third-order IFCs and provides a simpler and less expensive measure for the degree of anharmonicity in any material. 

The strong correlation between PBC$_{c=3}$ and the Grüneisen parameter arises from the fact that phonon frequencies inherently encapsulate how a lattice dynamically responds to structural perturbations. Unlike simple scalar quantities, such as average atomic mass, bond length, or volume, which reflect only static or average aspects of bonding, this descriptor captures the full vibrational spectrum through a frequency-weighted phonon density of states. This is physically significant because low-frequency phonon modes are more susceptible to volume changes and anharmonic effects. These modes correspond to collective atomic motions with large displacement amplitudes and long wavelengths, making them highly responsive to changes in interatomic distances and lattice volume. As a result, even smaller shifts in their frequencies as a result of expansion or compression can yield large Grüneisen parameters. Furthermore, these low-frequency modes can be thermally populated even at moderate temperatures; hence, their anharmonic contributions are amplified in the temperature-weighted Grüneisen parameter. By reflecting how vibrational states are distributed and weighted across the phonon spectrum, PBC$_{c=3}$ naturally emphasizes these thermally and anharmonically active modes. Therefore, PBC outperforms simpler quantities in capturing the effective anharmonic response of a material, as evident by its superior correlation with Grüneisen parameters.

\section{Conclusion}
In summary, the anharmonicity in materials originating from underlying atomic chemistry has been analyzed by considering a set of chalcopyrite compounds, selected through a high-throughput screening strategy. The anharmonic behavior of materials could be attributed to the mass of atoms, bonding framework, and nature of vibrations, collectively captured by the phonon density of states. Utilizing this advantage, we establish ``phonon band center (PBC)" as a potential descriptor to quantify the strength of anharmonicity in materials. PBC essentially characterizes the density of phonons that can contribute to enhancing the scattering channel and can be easily calculated by using only the second-order IFCs within the harmonic regime without the knowledge of higher-order potential terms. Additionally, PBC performs reasonably well in capturing the anharmonicity trends in other classes of materials with a wide range of variations in thermal conductivity. Our approach can be extended to develop high-throughput and ML models for the prediction of any thermal properties of class-independent, highly variable, comprehensive datasets.

% %\section*{\large{AUTHOR INFORMATION}}
% %\textbf{Corresponding Author}
% %$^\ast$Email:{~abhishek@iisc.ac.in}\\ 

\section*{Acknowledgements}
The authors thank the Materials Research Centre (MRC), Solid State and Structural Chemistry Unit (SSCU), and Indian Institute of Science for the computational facilities. A.S. and M.M. acknowledge the DST-INSPIRE fellowship [IF190068] and [IF160313]. The authors acknowledge the support from the Institute of Eminence (IoE) scheme of the Ministry of Human Resource Development (MHRD), Government of India.

\bibliography{manuscript}

\newpage
\onecolumngrid
\begin{center}
   \textbf{\Large Supplementary Material}
\end{center}
\renewcommand{\thefigure}{S\arabic{figure}}
\setcounter{figure}{0}
\section{Phonon Dispersions}
\begin{figure*}[!h]
    \centering
    \includegraphics[width=0.9\linewidth]{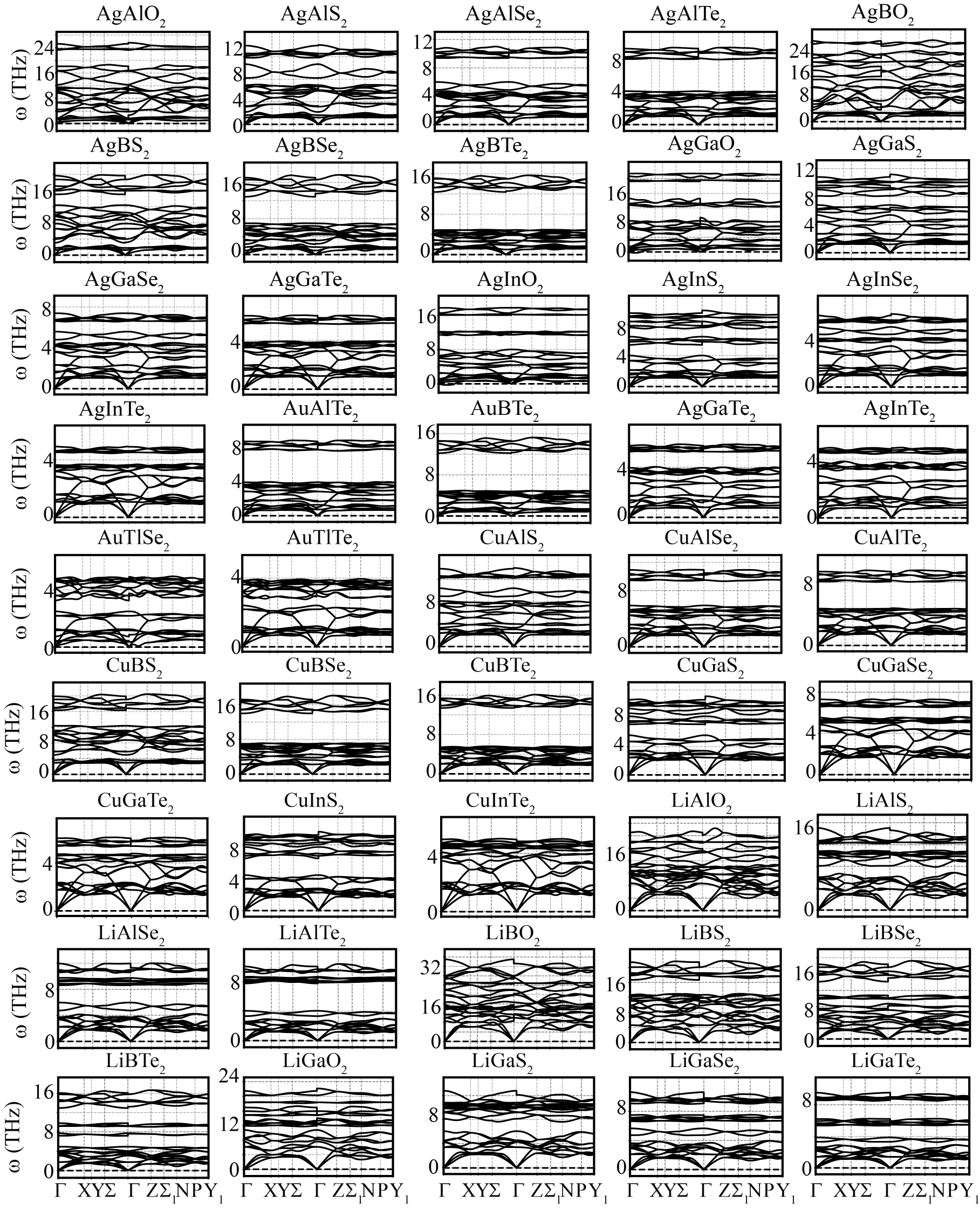}
    \caption{Phonon dispersions of chalcopyrties}
    \label{fig-ph1}
\end{figure*}

\begin{figure*}[!h]
    \centering
    \includegraphics[width=0.9\linewidth]{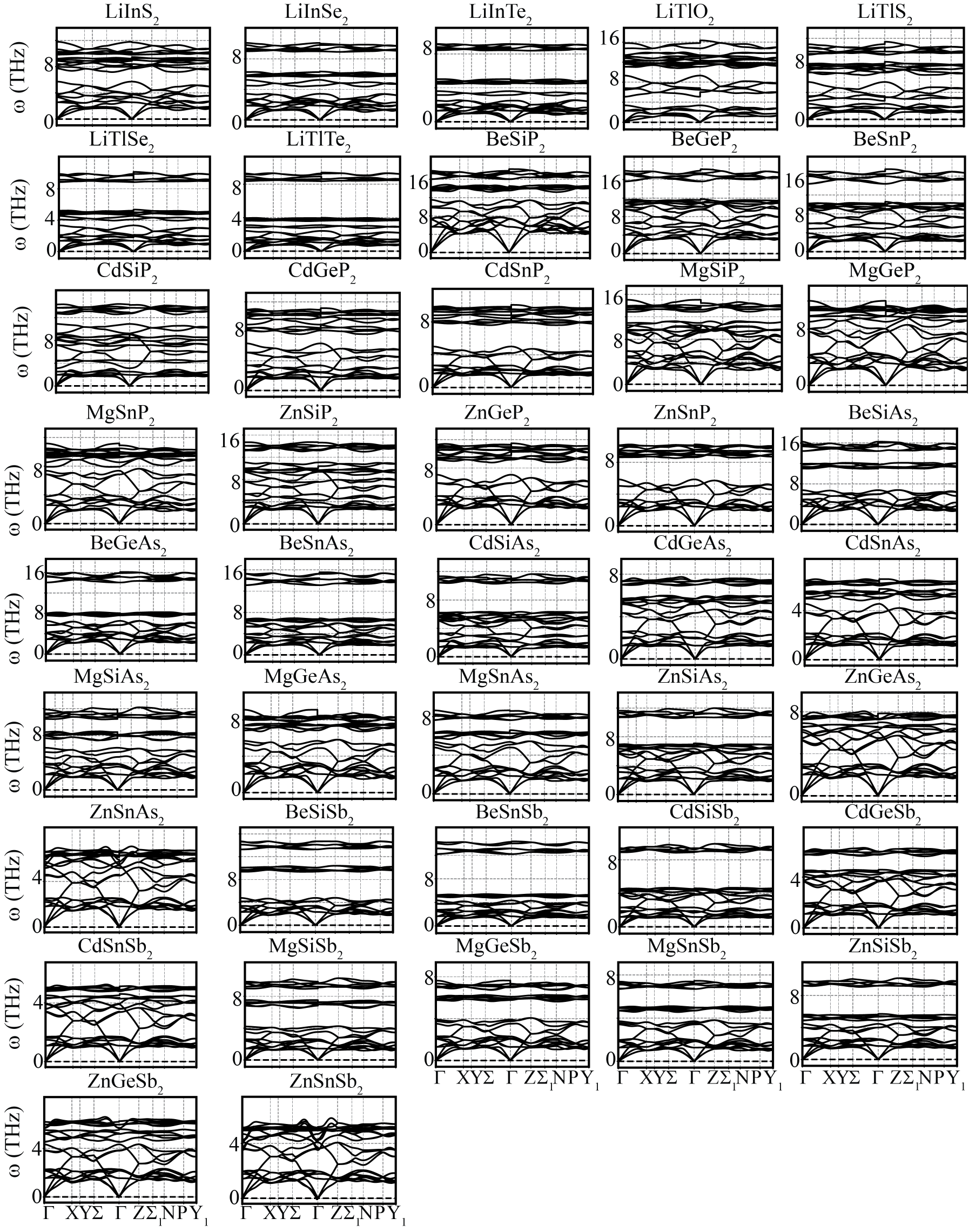}
    \caption{Phonon dispersions of chalcopyrties}
    \label{fig-ph2}
\end{figure*}

\newpage
\section{Gr$\ddot{u}$neisen parameter}
\begin{figure*}[!ht]
    \centering
    \includegraphics[width=0.95\linewidth]{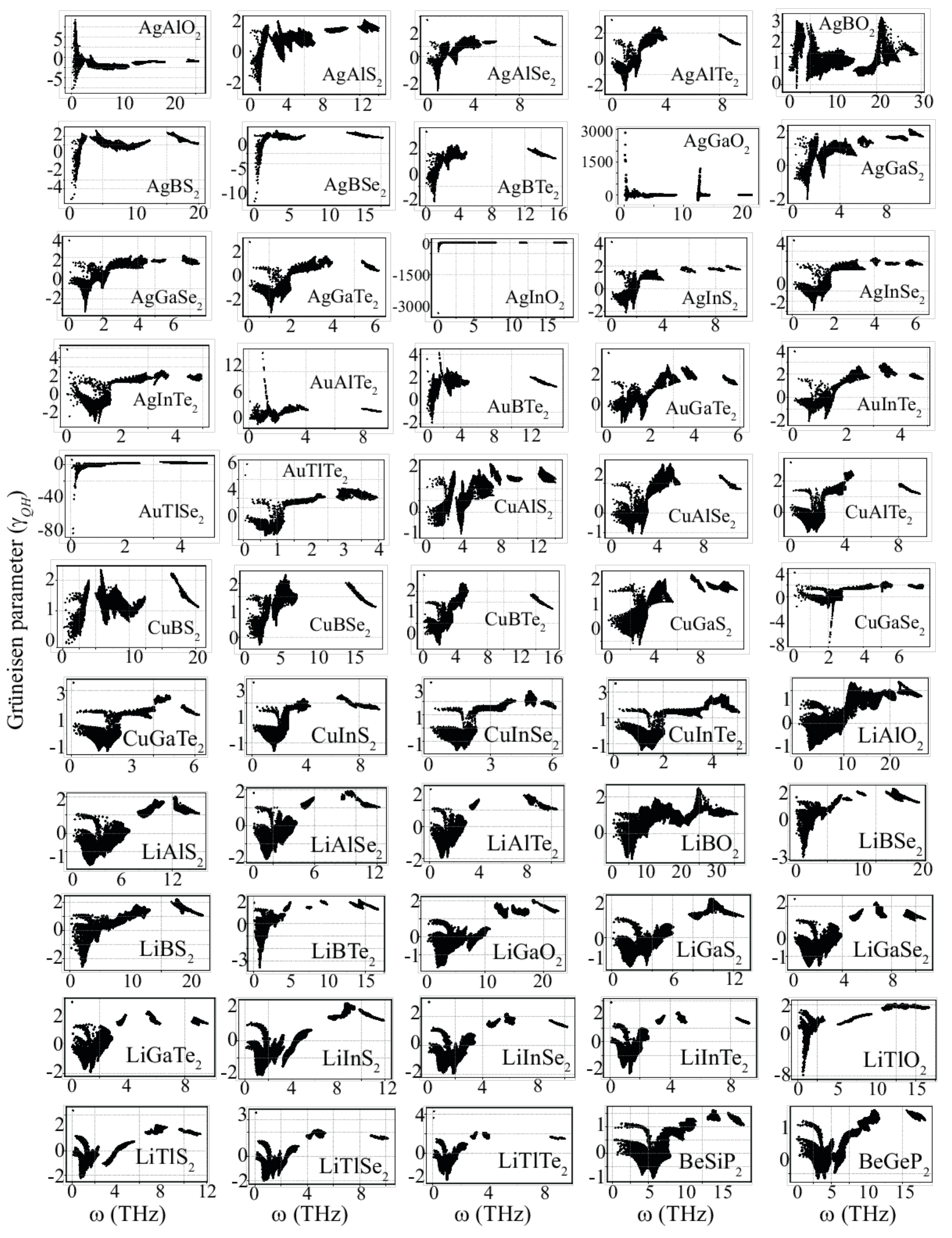}
    \caption{Gr$\ddot{u}$neisen parameter of chalcopyrties}
    \label{fig-gr1}
\end{figure*}

\newpage
\begin{figure*}[!h]
    \centering
    \includegraphics[width=0.95\linewidth]{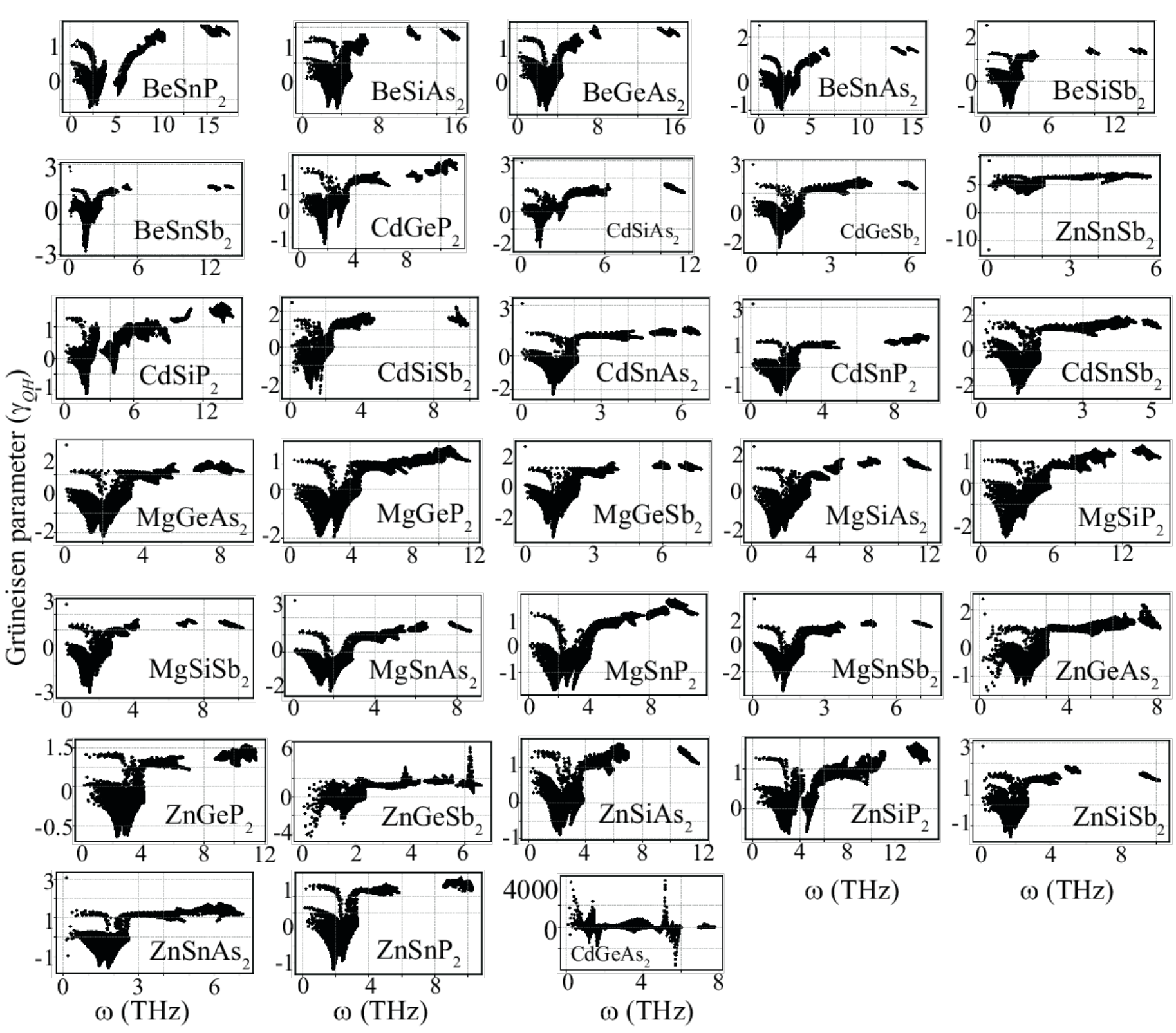}
    \caption{Gr$\ddot{u}$neisen parameter of chalcopyrties}
    \label{fig-gr2}
\end{figure*}

\newpage
\section{Phonon dispersion of three representative materials}
The phonon dispersion of three representative chalcopyrites, BeSiAs$_{2}$, BeSnSb$_{2}$, and AgInSe$_{2}$ corresponding to ultrahigh, medium, and ultralow range lattice thermal conductivity ($\kappa_l$), respectively.
\begin{figure*}[!h]
    \centering
    \includegraphics[width=1.0\linewidth]{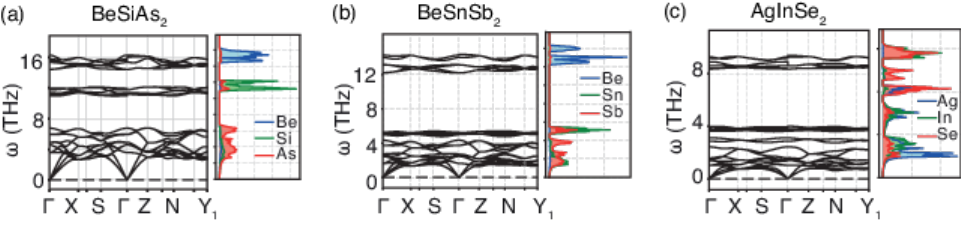}
    \caption{Phonon dispersion and corresponding atom-projected phonon density of states of (a) BeSiAs$_{2}$, (b) BeSnSb$_{2}$, and (c) AgInSe$_{2}$.}
    \label{fig-ph-dos}
\end{figure*}
The number of phonon states at lower frequencies increases as we move to ultrahigh to ultralow thermal conductivity materials.

\section{Potential energy surface}
The potential energy surface of any system represents the variation of potential energy with the change in the positions of atoms. In harmonic approximations, the potential energy is proportional to the square of the displacement from equilibrium, representing a parabola. Harmonic model assumes that bonds between the atoms act like perfect springs, which only holds true for small displacements. Anharmonicity refers to deviations from this ideal behavior, especially for large atomic displacements. Thus, for anharmonic systems potential energy surface deviates from a simple parabolic shape, becoming flatter or more complex at larger displacements.
\begin{figure*}[!h]
    \centering
    \includegraphics[width=1.0\linewidth]{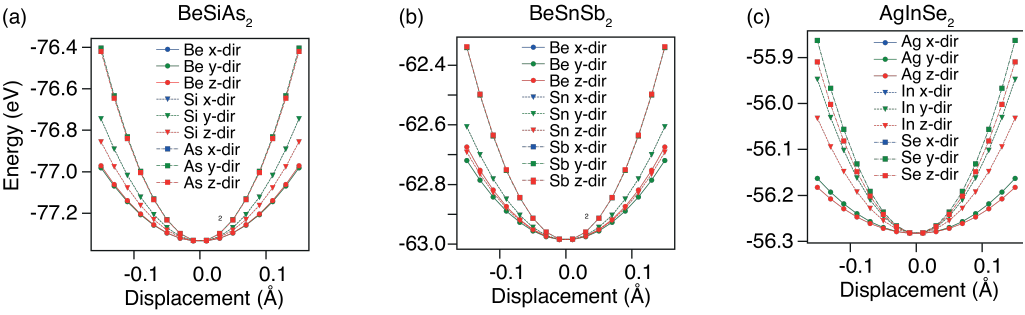}
    \caption{Potential Energy Surface of (a) BeSiAs$_{2}$, (b) BeSnSb$_{2}$, and (c) AgInSe$_{2}$}.
    \label{fig-pes}
\end{figure*}

\begin{table}[!htbp]
    \begin{center}
    \caption{Atomic contribution to total PBC for chalcopyrite (ABC$_{2}$) representative compounds.} 
    \label{table-PBC} 
\centering 
\begin{tabular}{|c|c|c|c|c|c|c|} 
\hline
\textbf{Material} & \textbf{PBC$_{A}$} & \textbf{PBC$_{B}$} & \textbf{PBC$_{C}$} & \textbf{PBC$_{all}$} & \textbf{$\gamma$}   \\ [1ex] 
\hline 
BeSiAs$_{2}$  & 14.39  & 10.15  & 5.33  & 8.71 &  3.90 \\ [1ex] \hline
BeSnSb$_{2}$  & 12.74  & 3.64  & 3.41  & 5.85 &  6.57 \\ [1ex] \hline
AgInSe$_{2}$  & 2.02  & 3.49  & 4.09  & 3.41 &  7.96 \\ [1ex] \hline
\end{tabular}
\end{center}
\end{table}
A flatter potential energy surface indicates that the energy of the system does not increase steeply with the displacement from equilibrium, which is a sign of anharmonic behavior. In a purely harmonic scenario, the energy is expected to increase rapidly (quadratically) with displacement. Anharmonic effects become more prominent when atomic displacements away from the minimum become flatter.

\section{Experimental validation of PBC}
\begin{figure*}[!h]
    \centering
    \includegraphics[width=1.0\linewidth]{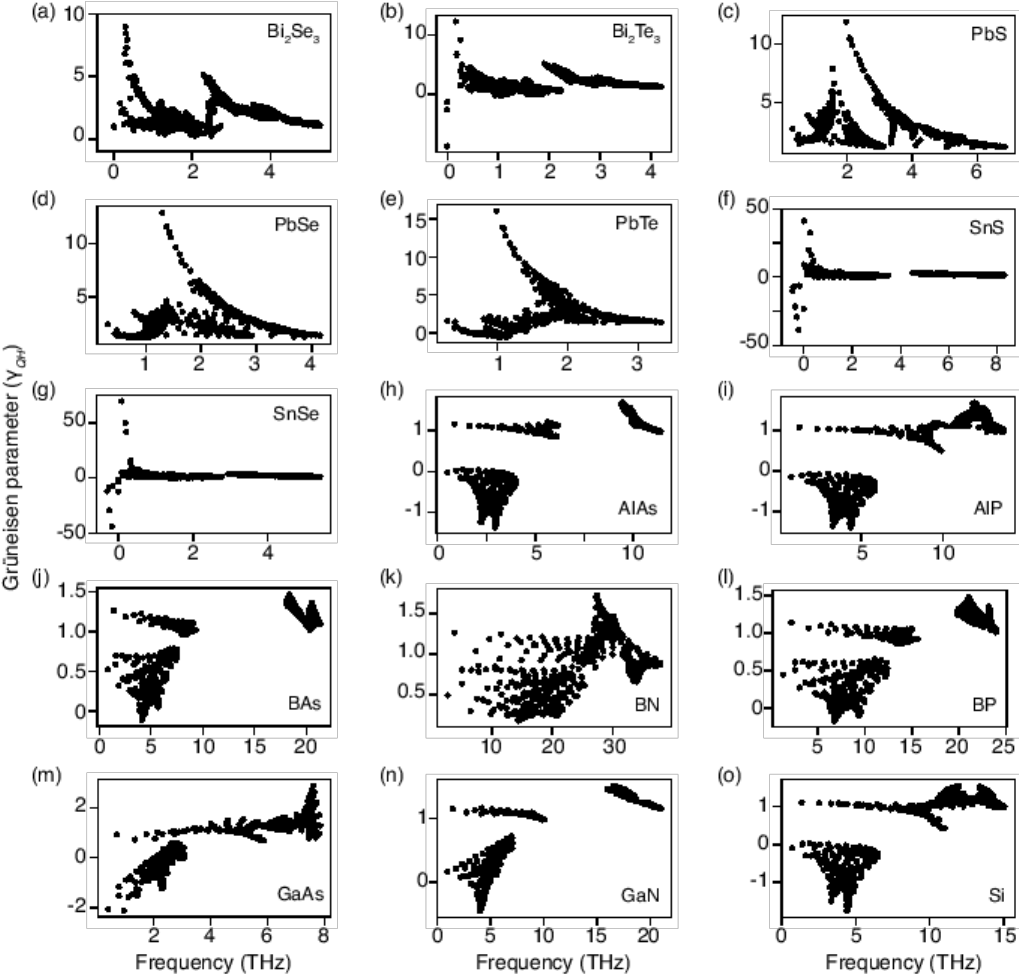}
    \caption{Gr$\ddot{u}$neisen parameter of experimentally available high and low lattice thermal conductivity systems (a) Bi$_2$Se$_3$, (b) Bi$_2$Te$_3$, (c) PbS, (d) PbSe, (e) PbTe, (f) SnS, (g) SnSe, (h) AlAs, (i) AlP, (j) BAs, (k) BN, (l) BP, (m) GaAs, (n) GaN, and (o) Si, respectively.}
    \label{expr-gr}
\end{figure*}

\newpage
\begin{table}[!htbp]
\caption{Expeimental lattice thermal conductivity ($\kappa_l$) of high and low heat-conducting materials at room temperature and corresponding PBC value calculated using \textit{first-principles}.} 
\centering 
\begin{tabular}{|c|c|c|c|} 
\hline
\textbf{Material} & \textbf{$\kappa_l$ (W/m-K)} & \textbf{PBC} & \textbf{References} \\ [1ex] 
\hline 
AlAs    &  98 & 6.83 & \cite{shinde2006high}  \\ [1ex] \hline
AlP    &  110 & 8.91 &  \cite{morelli2006high} \\ [1ex] \hline
% AlN    &  35 & 14.60 &   \\ [1ex] \hline
BAs    &  1300 & 12.86 &  \cite{kang2018experimental} \\ [1ex] \hline
BP    &  350 & 16.25 &  \cite{shinde2006high} \\ [1ex] \hline
BN    &  1200 & 25.61 &  \cite{karim1993characterization} \\ [1ex] \hline
GaAs    &  45 & 5.15 &  \cite{sze2021physics} \\ [1ex] \hline
% GaP    &  100 & 7.24 &  \cite{sze2021physics} \\ [1ex] \hline
GaN    &  220 & 12.13 &  \cite{jezowski2003thermal} \\ [1ex] \hline
Si    &  166 & 9.68 &  \cite{shinde2006high} \\ [1ex] \hline
Bi$_{2}$Se$_{3}$  &  2.5 & 2.89 &  \cite{fournier2018straightforward} \\ [1ex] \hline
Bi$_{2}$Te$_{3}$  &  1.62 & 2.30 &  \cite{bessas2012lattice} \\ [1ex] \hline
PbS  &  2.52 & 3.17 &  \cite{pei2012electrical} \\ [1ex] \hline
PbSe  & 1.83 & 2.28 &  \cite{pei2012electrical} \\ [1ex] \hline
PbTe  &  1.53 & 1.85 &  \cite{pei2012electrical} \\ [1ex] \hline
SnS  & 1.06  & 4.12 &  \cite{wu2023experimental} \\ [1ex] \hline
SnSe  & 0.62 & 2.92 &  \cite{zhao2014ultralow} \\ [1ex] \hline
% SnTe  &  2.88 & 1.82 &  \cite{banik2016origin} \\ [1ex] \hline
\end{tabular}
\label{table1} 
\end{table}
%\bibliography{supplementry}

\end{document}